\documentclass[twocolumn]{revtex4}
\usepackage{graphicx}
\begin{document}
\title{New self-gravito-acoustic mode  in degenerate quantum plasmas}
 \author{A. A. Mamun} \affiliation{Department of Physics,
Jahangirnagar University, Savar, Dhaka-1342, Bangladesh}
\begin{abstract}
The existence of a new  perturbation mode  [`self-gravito-acoustic  mode' (SGAM)] in cold self-gravitating degenerate quantum plasmas  (SGDQPs)  is theoretically predicted.  This new SGAM is developed in the perturbed  SGDQPs,  in which the compression is mainly provided by the self-gravitational pressure of the heavy particle species, and the rarefaction is mainly provided by the degenerate pressure of the light particle species.  The SGAM is a new  perturbation mode since it completely disappears if the degenerate pressure of the  light particle species is neglected.  The  prediction of this  new SGAM is applied in a white dwarf SGDQP.
\end{abstract}
\maketitle
The self-gravitating degenerate quantum plasma  medium \cite{Shapiro2004,Manfredi2005,Shukla2011,Haas2011}, which is common in many astrophysical compact objects (ACOs) like white dwarfs, neutron stars and black holes \cite{Shapiro2004},  significantly differs from other plasma media because of its extremely low temperate and extra-ordinarily high density.  
The ingredients of this self-gravitating degenerate quantum plasma  (SGDQP) medium  varies not only from one ACO to other, but also from one part to other parts of the same ACO. Thus, depending on ACO or its regions, the SGDQP  medium can be assumed to 
contain  non-inertial degenerate  light particle (viz. electron or/and positron or/and (non-zero
mass quark) species, and  inertial  degenerate  heavy (compared to
electron mass) particle (viz.  proton or/and neutron or/and $~^{4}_{2}$He or/and $~^{12}_{~6}$C or/and $~^{16}_{~8}$O),  $~^{56}_{26}$Fe, etc.) species.
The degeneracy of these particles arises due
to Heisenberg's uncertainty principle. This is due to the fact
that in a medium of extremely low temperature and extra-ordinarily high density,  the particles 
are located in an extra-ordinarily confined space, and that  the momenta of highly 
compressed particles are extremely uncertain. Thus, ts particles must move very fast on average, and gives rise 
to a very high pressure. This pressure is  known as `degenerate pressure',  which  depends on the degenerate 
particle number density, but not on  thermal temperature.  This generate pressure  is  given by \cite{Shapiro2004}  $P=3\hbar^2n^{5/3}/5m$ for non-relativistic limit. We note that the subscript $s$ and $j$ on $P$, $n$, and  $m$ are used for the species $s$ and $j$, respectively.  This outward degenerate pressure  is counter balanced by the inward self-gravitational pressure to form  ACOs  like white dwarfs, neutron stars, black holes, etc.  according to Chandrasekhar limit \cite{Shapiro2004}. 
  
The SGDQPM medium  in  ACOs  (particularly, in white dwarfs and neutron stars)  cannot be at equilibrium in general. This can be perturbed by many reasons (e. g. merging of two small ACOs \cite{Abbott2016}, splitting up a large ACO according to the Chandrasekhar limit \cite{Shapiro2004}, gravitational interaction
among nearby ACOs \cite{Shapiro2004},  etc.).  Once the SGDQP medium in any ACO is perturbed, a perturbation mode  [`self-gravito-acoustic  mode' (SGAM)]  developed by the compression (due to self-gravitational pressure on heavy particle species) and rarefaction (due to degenerate pressure on light particle species), and propagates through the medium.   The dynamics of this SGAM in such a SGDQP medium
is described by
\begin{eqnarray}
&& \frac{\partial \psi}{\partial
x}=-\frac{3}{2}\alpha_s\frac{\partial n_s^{\frac{2}{3}}}{\partial x},
\label{be1}\\
&&\frac{\partial n_j}{\partial t} +\frac{\partial}{\partial
x}(n_ju_j) = 0,
\label{be2}\\
&&\frac{\partial u_j}{\partial t} +u_j\frac{\partial
u_j}{\partial x}=-\frac{\partial \psi}{\partial
x}-\frac{3}{2}\beta_j\frac{\partial n_j^{\frac{2}{3}}}{\partial
x},
\label{be3}\\
&&\frac{{\partial}^2\psi}{\partial x^2}=\sum_s\sigma_s(n_s-1)+\sum_j \mu_j(n_j-1),
 \label{be4}
\end{eqnarray}
where $n_s$ ($n_j$) is the number density of the degenerate non-inertial
light (inertial heavy) particle species $s$ ($j$), and is normalized by
$n_{s0}$ ($n_{j0}$) in which $n_{s0}$ ($n_{j0}$) is
the equilibrium mass density of the degenerate particle species
$s$ ($j$);  $u_j$ is the degenerate fluid speed of the species
$j$, and is normalized by $C_q~(=\sqrt{\pi}\hbar
n_{e0}^{1/3}/m_q$, in which $q$ represents one of the species $j$  which have the maximum mass density (for example in white dwarfs it is $~^{12}_{~6}$C \cite{Koester1990} ;   
$\psi$ is  self-gravitational potential normalized by $C_q^2$; $\alpha_s=(m_q/m_s)^2(n_{s0}/n_{e0})^{2/3}$  (which represents the degeneracy effect of the non-inertial light particle species $s$)
$\beta_j=(m_q/m_j)^2(n_{j0}/n_{e0})^{2/3}$ (which represents the degeneracy effect of the inertial heavy particle species $j$),  $m_s$ ($m_j$) is the
mass of the degenerate particle species $s$ ($j$), and $n_{s0}$
($n_{j0}$) is the equilibrium number density of the degenerate
particle species $s$ ($j$);  $t$ is the time variable normalized by the time scale $\tau$,  which is the inverse of the Jeans frequency, $\omega_{Jq}=(4\pi G m_q n_{q0})^{1/2}$, in which  $G$ is the universal gravitational constant); $x$ is the space variable normalized by
the scale length $L_q$  (defined as $L_q=C_q/\omega_{Jq}$);   $\delta_s=m_sn_{s0}/m_qn_{q0}$, and 
$\mu_j=\omega_{Jj}/\omega_{Jq}$.  It should be noted here that the counter balance between  the outward degenerate pressure and the inward self-gravitational  pressure of non-inertial particle species $s$ gives rise to (\ref{be1}), which is however valid for the perturbation mode whose phase speed is much smaller than $C_{e}=\sqrt{\pi}\hbar
n_{e0}^{1/3}/m_e$.   

We first linearize  (\ref{be1})$-$(\ref{be4}) by assuming that $n_{s,j} =1+ \tilde{n}_{s,j}$, $u_j =0+\tilde{u}_j$, and  $\psi=0+\tilde{\psi}$, where $\tilde{n}_{s,j}$,  $\tilde{u}_j$, and  $\tilde{\psi}$  represent the perturbed part of  
$n_{s,j}$,  $u_j$,  and $\psi$.  We then assume that
all of these perturbed quantities  are directly proportional to
$\exp(-i\omega {t}+ikx)$, where $\omega$ is the  angular
frequency of the SGAM, and is normalized by $\omega_{Jh}$, and $k$
is the propagation constant, and is normalized by $L_q^{-1}$. This
assumption allows us to express the linear dispersion relation for
the SGAM as
\begin{eqnarray}
&&1+\sum_j\frac{\mu_j}{\omega^2-\beta_jk^2}-\sum_s\frac{\delta_s}{\alpha_sk^2}=0.
\label{dis}
\end{eqnarray}
This is the general dispersion relation for the SGPM propagating
in any SGDQP medium  containing arbitrary numbers of light non-inertial degenerate particle 
species $s$,  and  of  inertial heavy (compared to electron mass) degenerate particle  species $j$ . 
Thus, this general dispersion relation is valid for any SGDQP system.  However, to analyze this dispersion relation analytically, we first consider that all the heavy particle species $j$ are non-degenerate (i.e. $\beta_j=0$). This limiting case leads (\ref{dis}) to  
\begin{eqnarray}
&&\omega^2=k^2\frac{\sum_j\mu_j}{\sum_s\frac{\delta_s}{\alpha_s}-k^2}.
\label{dis-c1}
\end{eqnarray}
It is obvious from (\ref{dis-c1}) that when the non-inertial inertial particle species $s$ are assumed to be non-degenerate (i.e. 
$\alpha_s=0$),  (\ref{dis-c1}) reduces $\Omega=i\sqrt{\sum_j\omega_{Jj}^2}$, where $\Omega$ is the dimensional form of $\omega$. The latter means that for non-inertial light  light particle species (non-degenerate plasma medium), (\ref{dis-c1}) reduces to a well-known  purely growing unstable Jeans mode with the growth rate   
$\sqrt{\sum_j\omega_{Jj}^2}$.  However, for a degenerate plasma system (e.g. ACOs like white dwarfs and neutron stars), one cannot neglect $\alpha_s$ any way  \cite{Shapiro2004,Horn1991,Koester1990,Chandrasekhar1931}.  It is obvious  from (\ref{dis-c1})  that $\omega\rightarrow\infty$  at $k\rightarrow k_c$ (where $k_c=\sqrt{ \sum_s\delta_s/\alpha_s}$), and that  for $k\ll k_c$,  (\ref{dis-c1}) yields 
\begin{eqnarray}
&&\omega^2=k^2\left(\sum_j\mu_j\right)\left(\sum_s\frac{\alpha_s}{\delta_s}\right).
\label{dis-c2}
\end{eqnarray}
This represents the linear part of the dispersion relation for the predicted  new SGAM in  a degenerate plasma containing arbitrary number of degenerate light particle species $s$, and of non-degenerate inertial particle species 
$j$.  This is, in deed, represents a stable new  SGAM since the latter disappears when the effect of degeneracy all light particle species $s$ is neglected. This new stable mode exists for the wavelength satisfying the condition $k<k_c$. However, to examine the dispersion properties of this new mode for whole wavelength range, we apply  (\ref{dis-c1}) in a white dwarf SGDQP containing  non-inertial degenerate  electron species, and  inertial non-degenerate  nuclear ($~^{12}_{~6}$C) species (which has maximum mass density \cite{Koester1990}), and is denoted by $q$ ), and  inertial non-degenerate heavy  nuclear ($~^{56}_{26}$Fe) species (which  is the heaviest particle species, and is denoted by $h$).   

\begin{figure}[!h]
\centering
\includegraphics[width=8cm]{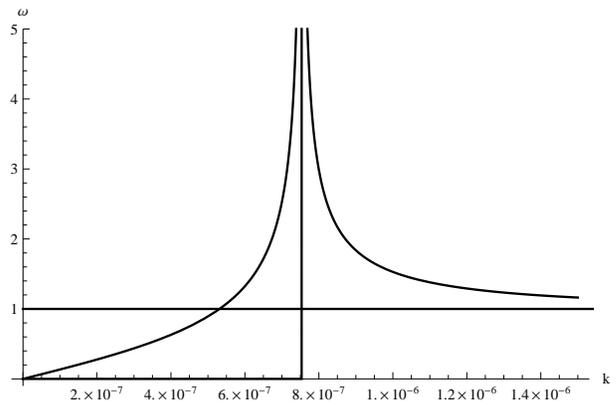}
\caption{Showing $\omega$ vs. $k$ curves
indicating  stable (before vertical line,  where $k=k_c$) and
unstable (after vertical line) regimes of the SGAM for the
typical parameters  corresponding to above mentioned white dwarf SGDQP.} 
\label{f1}
\end{figure}

To  compare among the effects of degeneracy in non-inertial electron species , and inertial  $~^{12}_{~6}$C and $~^{56}_{26}$Fe species, we have as estimated that   
$\alpha_e=4.85\times 10^{8}$,  $\beta_q=0.30$, and $\beta_h=2.43\times 10^{-4}$ for white dwarf SGDQP parameters (viz. $n_{e0}=10^{29}~{\rm cm^{-3}}$, $n_{q0}=0.99n_{e0}/Z_q$ and $n_{h0}=(n_{e0}-Z_qn_{q0})/Z_h$, in which $Z_q~(Z_h)$  is the number of protons in $q$ ($h$) species. Thus the assumption ($\beta_j=0$) used in (\ref{dis-c1} is quite valid  for a white dwarf SGDQP.  To see the nature of the dispersion curve in between $k\gg k_c$ and $k\ll k_c$, we now numerically solve  (\ref{dis-c1})  for typical parameters corresponding to a white dwarf SGDQP \cite{Shapiro2004,Koester1990,Vanderburg2015}, where  $m_q=12m_p$ and $m_h=56m_p$ (in which 
$m_p$ is the proton mass); $n_{e0}=10^{29}~{\rm cm^{-3}}$, $n_{q0}=0.995n_{e0}/Zq$  ($n_{q0}=n_{e0}/Zq$ means that no other nuclei of heavy elements are present).  The numerical results are shown in figure \ref{f1}.

To conclude, a new SGAM  in a cold SGDQP is identified for the first time.  The physics of this new SGAM is that if any SGDQP system is perturbed from its equilibrium state by its compression, the degenerate pressure  brings it back to its equilibrium state, but during this action it is expanded more than its equilibrium state according to Newtons first law of motion, and  again the self-gravitational pressure brings the system back to  its equilibrium state, but again during this action, it is compressed more than it equilibrium state. These compression and rarefaction of the medium continue, and consequently,  a low-frequency, long-wavelength new SGAM propagates through the medium. Though the result is applied in a particular SGDQP system like white dwarf, it can be applied to any SGDQP system since a SGDQP system containing an arbitrary number of degenerate, non-inertial, light  particle species $s$, and  an arbitrary number of degenerate, inertial, heavy  particle species  is considered.   Recent discovery \cite{Abbott2016} of the gravitational waves
\cite{Abbott2016} produced by merging of two
black holes leads us to expect that in near future the signatures
of similar or different kind of new wave/mode like SGAM
 in other astrophysical compact objects like white dwarfs and neutron stars
\cite{Shapiro2004} should be detected.

\end{document}